\documentclass{article}
\usepackage[latin9]{inputenc}
\usepackage{float}
\usepackage{textcomp}
\usepackage{amstext}
\usepackage{graphicx}
\usepackage[unicode=true]
 {hyperref}
\usepackage{breakurl}

\makeatletter

\floatstyle{ruled}
\newfloat{algorithm}{tbp}{loa}
\providecommand{\algorithmname}{Algorithm}
\floatname{algorithm}{\protect\algorithmname}

\usepackage{times}
\usepackage[accepted]{icml2016}
\usepackage{subfigure}

\usepackage{algorithm}
\usepackage{algorithmic}

\makeatother

\begin{document}
\twocolumn[
\icmltitle{Document Visualization using Topic Clouds}
\icmlauthor{Shaohua Li}{shaohua@gmail.com} 
\icmlauthor{Tat-Seng Chua}{dcscts@nus.edu.edu} 
\icmladdress{NExT Search Centre, National University of Singapore} 
] 
\begin{abstract}
Traditionally a document is visualized by a word cloud. Recently,
distributed representation methods for documents have been developed,
which map a document to a set of topic embeddings. Visualizing such
a representation is useful to present the semantics of a document
in higher granularity; it is also challenging, as there are multiple
topics, each containing multiple words. We propose to visualize a
set of topics using Topic Cloud, which is a pie chart consisting of
topic slices, where each slice contains important words in this topic.
To make important topics/words visually prominent, the sizes of topic
slices and word fonts are proportional to their importance in the
document. A topic cloud can help the user quickly evaluate the quality
of derived document representations. For NLP practitioners, It can
be used to qualitatively compare the topic quality of different document
representation algorithms, or to inspect how model parameters impact
the derived representations.
\end{abstract}

\section{Introduction}

Word clouds (also known as ``tag clouds'') are a conventional way
to visually represent the words in a document \cite{headin}. Typically
the font size of a word is proportional to its importance\footnote{The importance of a word is usually defined as a function of its frequency,
or the TF-IDF score as in \cite{tfidf}.} in the document. Figure 1 presents a frequency-weighted word cloud
in a typical style, generated from a news report about a pharmaceutical
company acquisition\footnote{http://www.nytimes.com/2015/09/21/business/a-huge-overnight-increase-in-a-drugs-price-raises-protests.html}.
The colors of words are randomly selected from a palette, without
semantic indications.

One apparent problem of the word cloud is that, as the complexity
of the document increases, it soon becomes difficult to read. For
instance, Figure 1 only contains 60 words, but a viewer will probably
only notice the few largest words, and could not form a ``big picture''
of the document, as the semantic transition across words are random
and abrupt. \cite{clustering} proposed to cluster words according
to their semantic relatedness, and draw differnt clusters in different
lines. This alleviates the unorganized nature of the word cloud to
certain extent.

\begin{figure}
\noindent \centering{}\includegraphics[scale=0.3]{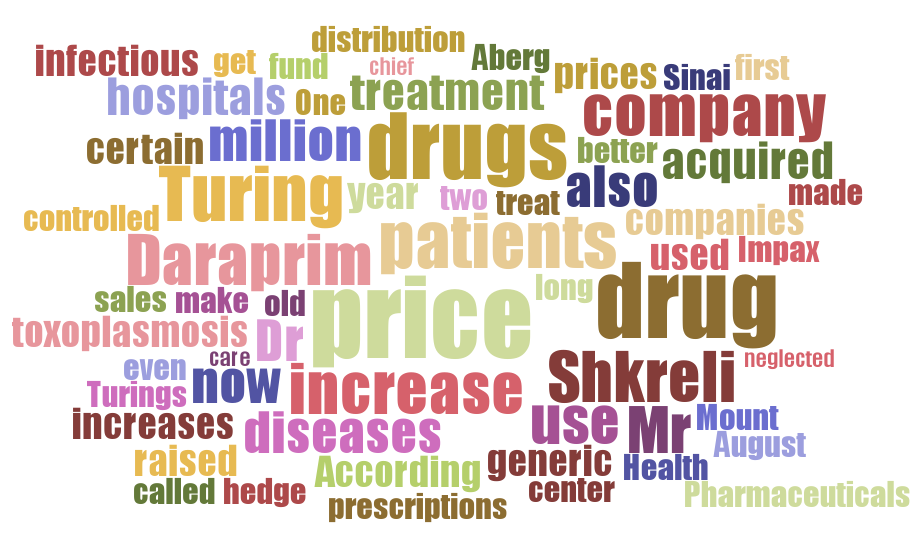}\caption{A typical word cloud generated from a news report, containing the
60 most frequent words (stop-words are removed).}
\end{figure}

The advent of distributed representations (``embeddings'') of words
and text has led to an evolution of Natural Language Processing \cite{nlpscratch,word2vec}.
Embedding methods map words and text into continuous feature vectors
in a low-dimensional space, making them easy to process by downstream
machine learning algorithms. Recently, a few methods have been proposed
to map documents into a set of embeddings \cite{doc2vec,liu,gaussianLDA,vmf,topicvec}.
Most of these works derive embeddings that are \emph{topical}, i.e.,
each embedding defines a topic (a distribution of words) of the document.
Compared to conventional topic models \cite{lda}, these methods derive
more coherent topics by exploiting semantic relatedness encoded in
pretrained word embeddings; moreover, some of them, e.g. \cite{topicvec},
are able to derive topic embeddings based on only one document.

Topic embeddings, along with the corresponding topic proportions,
represent the semantics of a document in a refined granularity. Visualizing
the topic embddings, or the corresponding topics, can help users quickly
perceive the main concepts in a document. However there are multiple
topics, each containing multiple words, and words/topics differ in
their prominence. It is challenging to represent the topics in a form
that is both visually organized, and also manifests the different
prominence of words/topics. To this end, we propose Topic Cloud, which
is a pie chart consisting of topic slices, where each slice contains
important words in this topic. The relative prominence of words/topics
are made explicit by drawing the words/topics in sizes that are proportional
to their importance in the document. Figure 2 provides an example
of topic clouds.

\begin{figure}
\noindent \centering{}\includegraphics[scale=0.2]{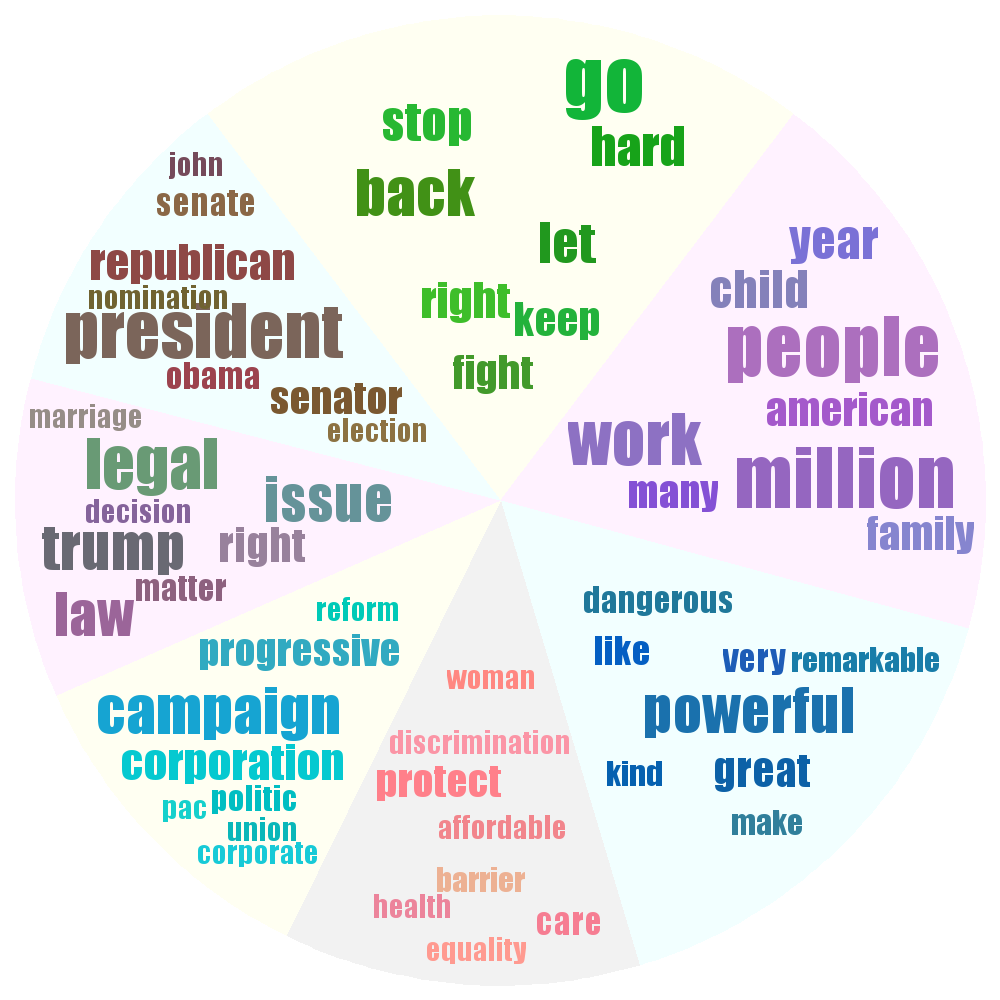}\caption{A topic cloud generated from one of Hillary Clinton's presidential
campaign speech.}
\end{figure}

The topic cloud is an easily recognizable visualization of the topical
representation of a document\footnote{Multiple documents can be used to derive one set of topic embeddings,
hence the topic cloud can be adopted to visualize multiple documents
as well.}. It helps the user quickly perceive the main concepts in a document.
In addition, it also makes it easy for the user to evaluate the quality
of derived representations of documents. For NLP practitioners, It
can be used to qualitatively compare the topic quality of different
document representation algorithms, or to inspect how model parameters
impact the derived representations.

The source code of our Topic Cloud implementation is available at
\href{https://github.com/askerlee/topiccloud}{https://github.com/askerlee/topiccloud}.

\section{The Topic Cloud Algorithm}

\begin{algorithm}
\caption{Topic cloud generation.}

\textbf{Input}: $K$ topics $T=\{t_{1},\cdots,t_{K}\}$, where $t_{k}=p_{k},(w_{k1},q_{k1};\cdots,w_{km},q_{km})$;
exponential scaling coefficient $\beta$, thresholds of topic proportion
ratio $\mu$ and word importance $\sigma$, maximal and minimal font
sizes $f_{\textrm{max}}$,$f_{\textrm{min}}$; \medskip{}

Draw a circle of radius $r$ as the canvas;

Sort all topics in descending order of their proportions $p_{k}$;

Remove all topics $t_{k}$ satisfying $p_{k}<p_{1}/\mu$;

Normalize $p_{k}$ as $p_{k}'=\frac{p_{k}^{\beta}}{\sum_{j}p_{j}^{\beta}};$

$q_{\textrm{max}}=\max_{k,i}q_{ki}$;

$a_{0}=270-180\cdot p_{k}'$;\smallskip{}

\textbf{for} $t_{k}$ in $T$ \textbf{do}

\quad{}%
\noindent\begin{minipage}[t]{1\columnwidth}%
Allocate $t_{k}$ a pie slice with angles in $[a_{0},a_{0}+360\cdot p_{k}']$;

Draw the slice $s_{k}$ with background color $c_{1+i\textrm{ mod }L_{1}}\in\{c_{1},\cdots,c_{L_{1}}\}$,
a predefined palette;

Set the base word color in $t_{k}$ as $\textrm{bg}_{k}=d_{1+k\textrm{ mod }L_{2}}\in\{d_{1},\cdots,d_{L_{2}}\}$,
another predefined palette;

Sort $w_{k1},\cdots,w_{km}$ in descending order of $q_{ki}$;

Remove all words satisfying $q_{ki}<\sigma$;\smallskip{}

\textbf{for} $w_{ki}$ in $t_{k}$ \textbf{do}

\quad{}%
\noindent\begin{minipage}[t]{1\columnwidth}%
Set the font size of $w_{ki}$ as $f_{ki}=f_{\textrm{max}}\cdot(q_{ki}/q_{\textrm{max}})^{\beta}$;

$f_{ki}=\max(f_{ki},f_{\textrm{min}})$;

Compute the bounding box $B_{ki}$ of $w_{ki}$ in font size $f_{ki}$;

\textbf{repeat}

\quad{}%
\noindent\begin{minipage}[t]{1\columnwidth}%
Find all points $S=\{(x_{j},y_{j})\}$ within slice $s_{k}$ that
can be used as the upperleft corner of $B_{ki}$, i.e. all points
within $B_{ki}$ is unoccupied;

\textbf{if} $S=\phi$ \textbf{then}

\quad{}$f_{ki}=f_{ki}-1$

\textbf{end if}%
\end{minipage}

\smallskip{}
 \textbf{until} $S\ne\phi$

Randomly pick $(x,y)\in S$;

Let $\textrm{bg}_{k}=\textrm{rgb}(r_{k},g_{k},b_{k})$. Randomly perturb
$r_{k},g_{k},b_{k}$ by a random integer in $[-\epsilon,\epsilon]$,
and get $\textrm{rgb}(r_{k}',g_{k}',b_{k}')$ as the color of $w_{ki}$;

Draw $w_{ki}$ in $B_{ki}$ at $(x,y)$;

Mark all points in $B_{ki}$ as being occupied;%
\end{minipage}

\smallskip{}
 \textbf{end for}

$a_{0}=a_{0}+360\cdot p_{k}'$;%
\end{minipage}

\textbf{end for} 
\end{algorithm}
The topic cloud generation algorithm receives a set of topics as the
input, where each topic $t_{k}$ is in the form of $p_{k},(w_{k1},q_{k1};\cdots,w_{km},q_{km})$.
Here $p_{k}$ is the proportion of $t_{k}$ in the represented document,
$m$ is a pre-specified word number threshold, $w_{ki}$ is a word
belonging to $t_{k}$, and $q_{ki}$ is its relative importance. As
a preprocessing step, we lemmatize all words in each topic. If two
words $w_{ki},w_{kj}$ are lemmatized into the same word $w_{ki}'$,
then their importance is combined as $q_{ki}'=q_{ki}+q_{kj}.$

The topic cloud generation algorithm is straightforward, as described
in Algorithm 1. For convenience of computation, the 90\textdegree ~
angle is defined at the center bottom of the canvas. We start putting
topics clockwise from the center top, i.e. around the 270\textdegree ~
angle.

\section{Example Applications}

Traditionally, the quality of derived topical representations is usually
measured by the model perplexity, or by the Pointwise Mutual Information
(PMI) score of the words in the topic against a golden standard. But
the perplexity is not intuitive, and the perplexity between different
methods may be incomparable. On the other hand, the PMI score is costly
to compute.

When we only want to informally evaluate the quality of derived representations,
we could resort to qualitative analysis. Qualitatively, the topical
representations can be measured in two aspects: 1) whether the words
in each major topic is semantically coherent; 2) whether the proportions
of topics comply with the following intuitions: the topics in a document
are usually sparse, i.e., only a few major topics take most of the
proportions, and other topics have minor proportions; on the other
hand, topic proportions are usually somewhat uneven (gradually decreasing).

The topic cloud can be used to qualitatively evaluate these two aspects
of derived representations. Here we present an application of comparing
the performance of two topic embedding methods, followed by an application
of tuning a parameter of a topic embedding method.

\subsection{Comparison of Topical Representations by K-Means and TopicVec}

In this example, the two compared methods are a simple k-means clustering
algorithm on the word embeddings and TopicVec \cite{topicvec}. The
topic numbers of both methods were set to 10. As the cosine similarity
measures the semantic relatedness between embeddings, the metric of
k-means was specified as the cosine distance. Before performing k-means,
the embedding vectors were normalized. The input document was the
pharmaceutical company acquisition news report, the same input of
Figure 1.

Figure 3 and 4 present the topic clouds derived by k-means and TopicVec,
respectively. For k-means, each cluster was a topic, and the cluster
centroid (the average embedding in a cluster) was used as the topic
embedding. The topic proportion was defined as the proportion of words
in this cluster. Constrained by the limited circle area, only the
6 biggest topics were shown in each topic cloud. In the following,
we refer to the center top topic slice as \emph{the first topic},
which is always the biggest slice.

One can quickly see that in Figure 3, the first two topics (clockwise
counted) are coherent, and the remaining 4 topics become increasingly
noisy. In contrast, the topics in Figure 4 are generally coherent,
with very few noisy words.

By comparing Figure 3 and 4, we can see that topics produced by k-means
are more even, with similar sizes; while the topics produced by TopicVec
are more disproportionate. The latter agrees better with the intuition
of topic sparsity.

In sum, with the help of topic clouds, one can quickly learn that
TopicVec derives better topical representations than k-means, both
in topic coherence and topic proportions.

\begin{figure}
\noindent \centering{}%
\noindent\begin{minipage}[t]{1\columnwidth}%
\noindent \begin{center}
\includegraphics[scale=0.2]{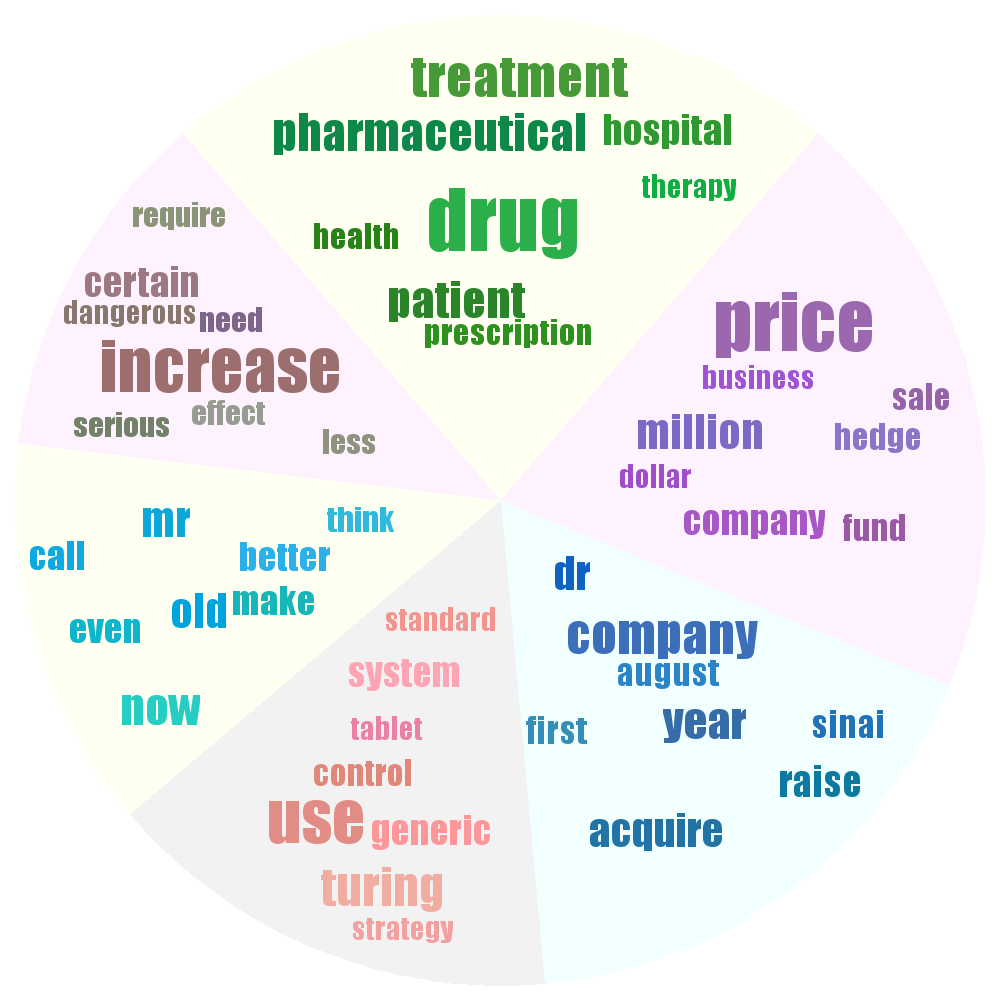} 
\par\end{center}
\caption{The topic cloud derived by k-means.}
\end{minipage}\hfill{}%
\noindent\begin{minipage}[t]{1\columnwidth}%
\noindent \begin{center}
\includegraphics[scale=0.2]{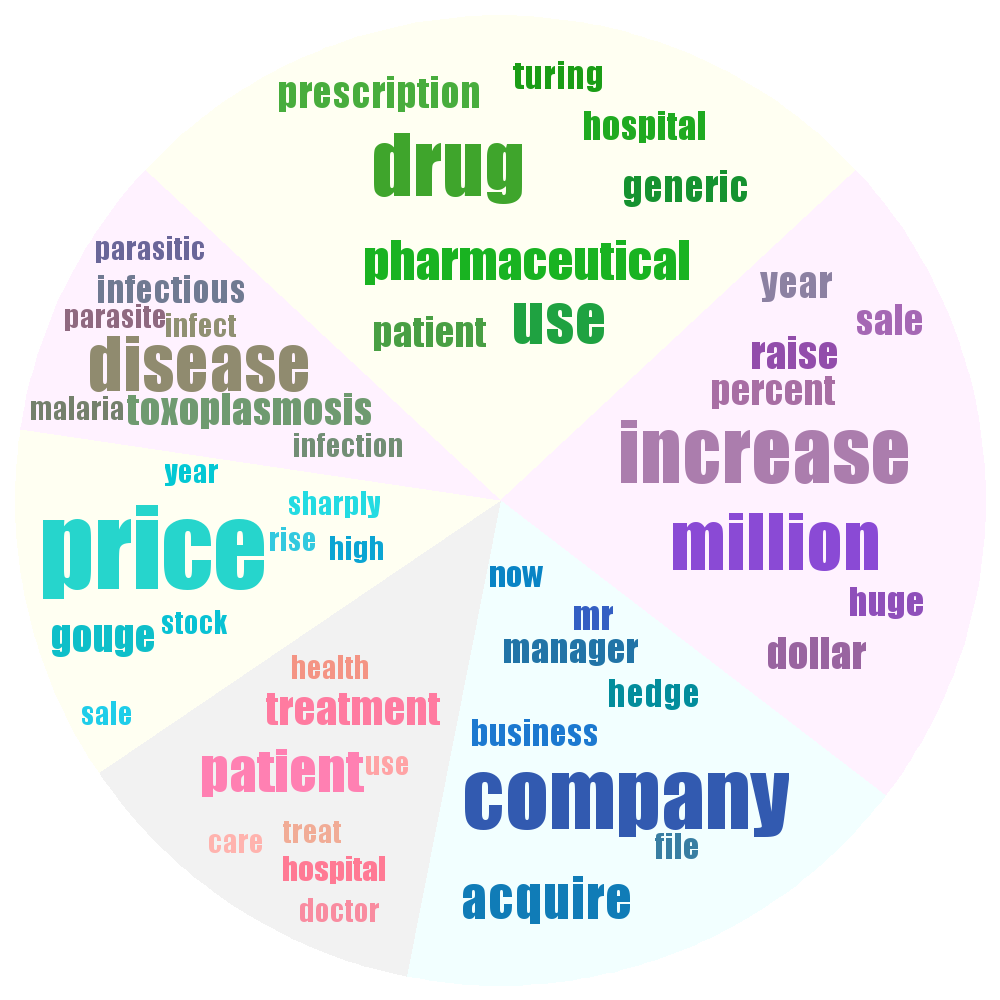} 
\par\end{center}
\caption{The topic cloud derived by TopicVec.}
\end{minipage}
\end{figure}

\subsection{Impact of Parameters on Topics by TopicVec}

In this example, we tune an important parameter of TopicVec, i.e.,
the maximal magnitude $\gamma$ of topic embeddings.

Figure 5 and 6 present the topic clouds derived by TopicVec on the
accepted paper list of ICML 2016. $\gamma$ was set to 3 and 5, respectively.
In Figure 5, one can quickly find out that all topics except the first
one are highly similar, and all topics have similar proportions. In
contrast, in Figure 6, words are clustered into different coherent
topics, and the topic proportions gradually decrease clockwise. The
two topic clouds reveal that $3$ is a poor setting of $\gamma$,
and 5 is reasonable.

\begin{figure}
\noindent \centering{}%
\noindent\begin{minipage}[t]{1\columnwidth}%
\noindent \begin{center}
\includegraphics[scale=0.2]{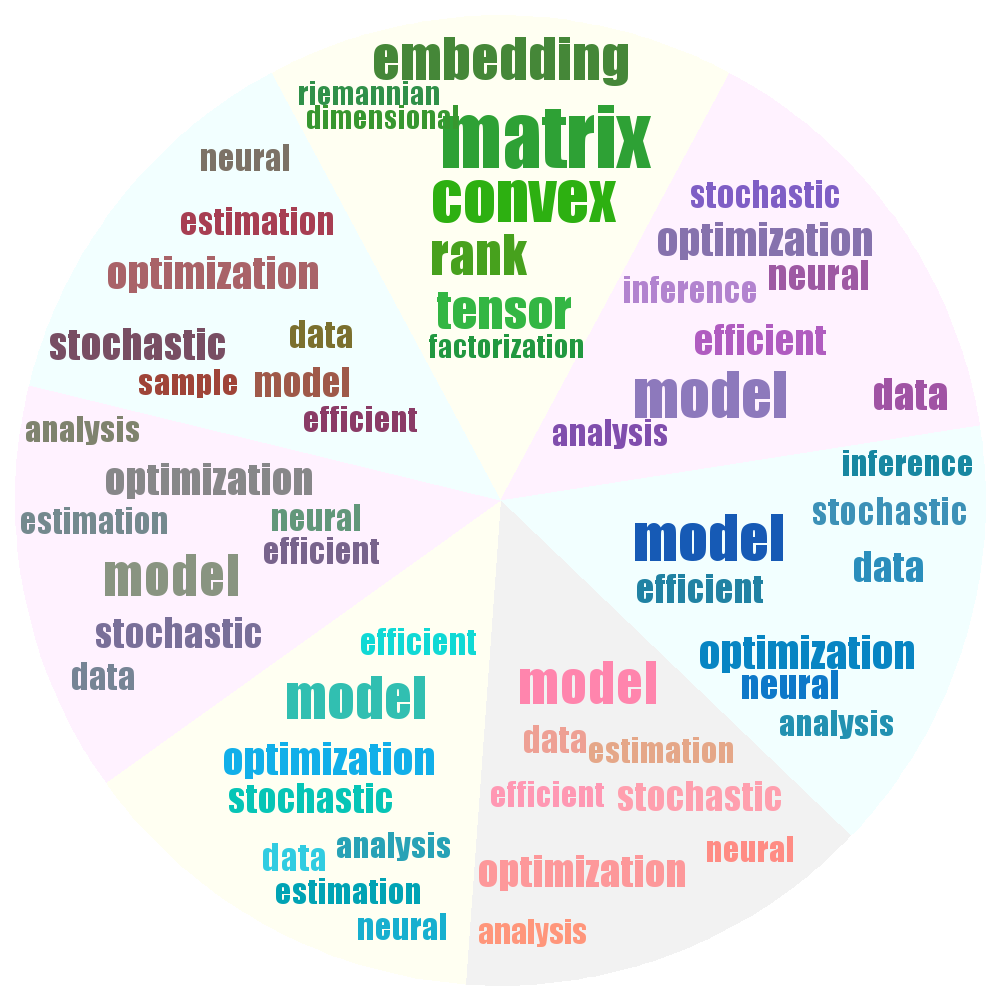} 
\par\end{center}
\caption{The topic cloud derived by TopicVec with $\gamma=3$, on the accepted
paper list of ICML'16.}
\end{minipage}\hfill{}%
\noindent\begin{minipage}[t]{1\columnwidth}%
\noindent \begin{center}
\includegraphics[scale=0.2]{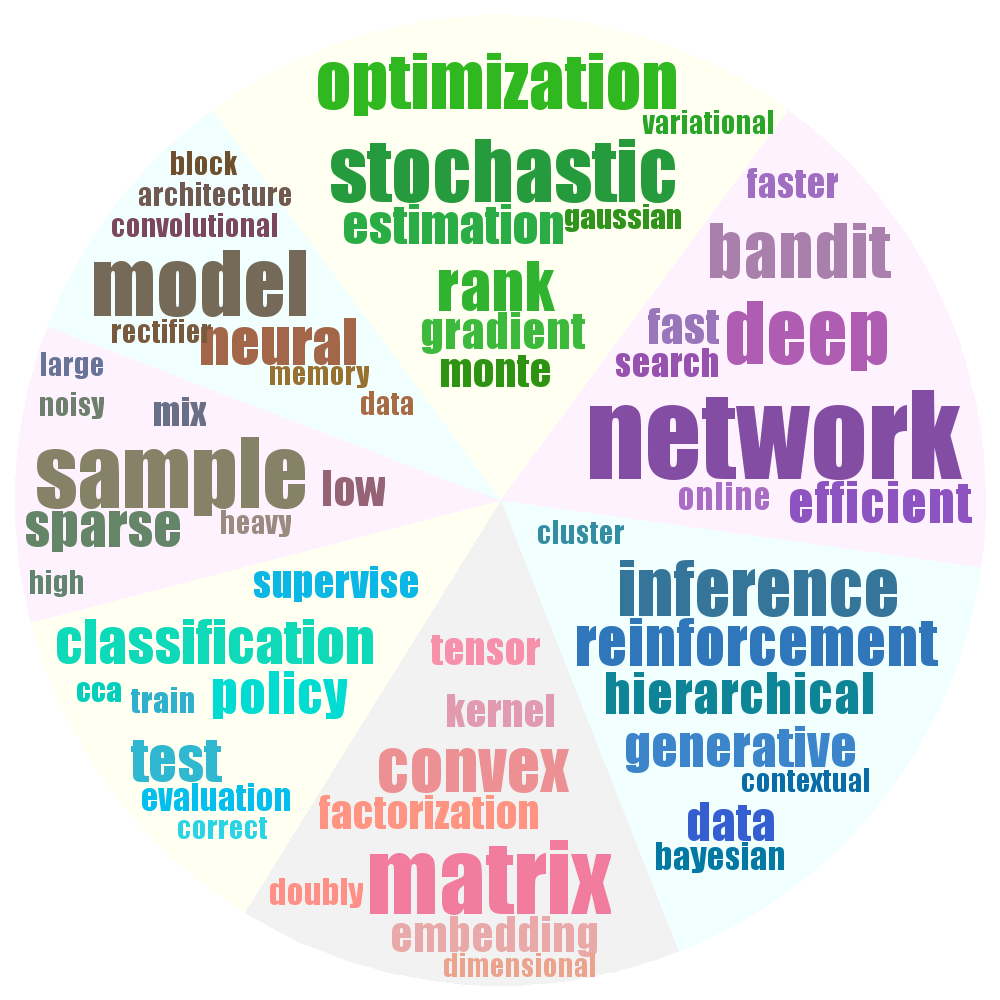} 
\par\end{center}
\caption{The topic cloud derived by TopicVec with $\gamma=5$, on the accepted
paper list of ICML'16.}
\end{minipage}
\end{figure}

\section{Future Work}

Our method of Topic Cloud generation is still preliminary. One deficiency
of the present method is that, the words within each topic are placed
randomly, without considering their semantic relatedness. It would
be easier for human to perceive if, within each topic, words are arranged
according to their semantic relatedness, i.e. more relevant words
are put more closely. Distance preserving dimension reduction methods,
such as t-SNE \cite{tsne} (extension is needed to incorporate boundary
and word size constraints), could be adopted to perform a projection
from word embeddings within a topic to a pie slice. With such a technique,
the drawn topic cloud will be visually more coherent, allowing users
to more quickly recognize the concepts in each topic.

 \bibliography{topiccloud}
 \bibliographystyle{icml2016} 
\end{document}